\documentclass[preprint]{revtex4}
\usepackage{times,xspace}
\usepackage{amsbsy,amssymb,amsmath,bm,mathptmx}
\usepackage{graphicx,color,epsfig,rotate}
\usepackage{graphics}
\usepackage{amsmath}

\begin{document}

\title{Phenomenology of altermagnets}

\author{Maxim Mostovoy}
\thanks
{Corresponding author} 
\email{m.mostovoy@rug.nl}
\affiliation{Zernike Institute for Advanced Materials, University of Groningen, Nijenborgh 3, 9747 AG Groningen,  The Netherlands}

\date{\today}

\begin{abstract}
Altermagnets have recently emerged as a new class of magnetic materials sharing properties of both antiferromagnets and ferromagnets. Despite very small net magnetization, they show phenomena usually associated with ferromagnetism, such as the Faraday, Kerr and Anomalous Hall effects, resulting from the relativistic spin-orbit coupling, as well as the spin splitting of electron bands and Spin Hall Effect of non-relativistic origin.  Spin space groups and magnetic multipoles are used to explain symmetry properties of altermagnets. Here, I show that the conventional phenomenological description in terms of a vector antiferromagnetic order parameter can be applied to all effects observed in altermagnets with collinear and non-collinear spin orders. I also discuss non-relativistic effects in non-altermagnets.	 			
\end{abstract}

\maketitle


Ferromagnetism has been known to mankind for a very long time and the presence of spontaneously ordered magnetic moments can be detected by a variety of experimental techniques, e.g. by measuring the rotation of light polarization or Anomalous Hall Effect (AHE). 
%
On the other hand, antiferromagnetism was understood only some $100$ years ago.

Surprisingly, some antiferromagnets, now called altermagnets, show the AHE \cite{Nakatsuji2015,Nayak2016}, anomalous Nernst effect \cite{Guo2017,Ikhlas2017,Li2017} and magneto-optical Kerr effect \cite{Higo2018,Wu2020} with magnitudes comparable to those in ferromagnets, even though they induce very weak magnetic fields.  
Similarly to the exchange splitting in ferromagnets, electron bands of itinerant altermagnets are spin-polarized in the absence of spin-orbit coupling  \cite{Hayami2019,Yuan2020,Smejkal2020}.
This spin splitting, observed by the Angle-resolved Photoemission Spectroscopy (ARPES) \cite{Bai2022,Ding2024,Reimers2024,Lee2024,Krempasky2024,Osumi2024}, was predicted to result in an efficient charge-to-spin conversion \cite{Naka2019,Gonzales2021}.

%
Altermagnets were pronounced to be a new class of magnetic materials sharing properties of both `conventional antiferromagnets' and ferromagnets \cite{Smejkal2022a}.
Not all researchers agree, arguing that the Faraday and Anomalous Hall effects in antiferromagnets have a long history \cite{Krichevtsov1981,Zenkov1989,Vlasov1980,Kimel2024}. 

In the author's view both camps have a point.
%
Symmetry-wise, altermagnets belong to the well-understood group of antiferromagnets showing a weak ferromagnetic moment (WFM) or acquiring one under strain (piezomagnetism). 
On the other hand, earlier research on magneto-optical and transport effects in antiferromagnets was focused on Mott insulators, such as YFeO$_3$ \cite{Krichevtsov1981,Zenkov1989} and Fe$_2$O$_3$ \cite{Vlasov1980}, whereas recent studies brought into the spotlight conducting antiferromagnets with high transition temperatures that are potentially useful for spintronics \cite{Rimmler2025}.
%

It was argued that since magnetic groups make no distinction between relativistic and non-relativistic effects, symmetry analysis of altermagnets requires more sophisticated methods, such as spin space groups \cite{Smejkal2022b,Liu2022} or magnetic multipoles \cite{Suzuki2017,HayamiYanagi2020}. 
The aim of this paper is to show that the conventional approach using the N\'eel order parameter and crystal symmetries can describe all physical phenomena observed in altermagnets and differentiate between relativistic and non-relativistic contributions. 
%
%
This description can be incorporated into Landau theory and is particularly useful when spins change direction under applied electromagnetic fields, strains or stresses.

%
%


Two-sublattice (nearly) collinear antiferromagnets are described by the N\'eel vector, $\mathbf{l} = \frac{1}{2}\left(\mathbf{m}_1 - \mathbf{m}_2\right)$, $\mathbf{m}_{1,2}$ being the sublattice magnetizations. 
Detection of antiferromagnetism by macroscopic effects requires a coupling to uniform electromagnetic fields, currents or strains, for which $\mathbf{l}$ must be invariant under crystal translations.
Furthermore, to induce a WFM, an antiferromagnetic (AFM) ordering must preserve inversion symmetry, i.e. $\mathbf{l}$ must be even under inversion, $I$. 
Invariance under translations and inversion is a characteristic property of altermagnets.
AFM states odd under $I$ show the linear magnetoelectric effect (LME) -- an electric polarization proportional to an applied magnetic field and a magnetization induced by an applied electric field \cite{Landavshitz}. 
In centrosymmetric crystals, weak ferromagnetism and LME are mutually exclusive.

The gyration vector governing the rotation of light polarization, $\mathbf{g}(\omega)$, $\omega$ being the light frequency, is dual to the antisymmetric part of permitivity tensor and, therefore,  transforms as the axial  magnetization vector, $\mathbf{M}$, under rotations and inversion of the crystal lattice  \cite{Landavshitz}.
By virtue of Onsager reciprocity relations, $\mathbf{g}(\omega)$ changes sign under time reversal, $T$, together with $\mathbf{M}$ and $\mathbf{l}$. 
The same holds for the Hall vector, $\mathbf{h}\propto\omega \mathbf{g}(\omega)$ in the zero frequency limit, dual to the antisymmetric part of conductivity tensor. 
The fact that  $\mathbf{M}, \mathbf{h}$ and  $\mathbf{g}$ have the same symmetry does not imply that the Faraday, Kerr and Anomalous Hall effects result from the WFM, but if some component of $\mathbf{M}$ is zero by symmetry, the same holds for $\mathbf{h}$ and  $\mathbf{g}$.

%
%

The WFM is of crucial importance for altermagnets, since it allows for selection of AFM domains.
An effective and elegant aproach to weak ferromagnetism in collinear antiferromagnets has been  developed by Turov \cite{Turov2010}. 
Since translation and inversion symmetries are not broken in weak ferromagnets, it suffices to consider rotations of the crystal lattice. 
Each rotation $R$ is characterized by a sign factor, $\sigma_R$, equal $-1$, if $R$ interchanges the two  magnetic sublattices and $+1$, otherwise. 
The N\'eel vector, $\mathbf{l}$, transforms under $R$ as a $T$-odd axial vector, like the magnetic field $\mathbf{H}$, but gets an additional sign factor $\sigma_R$. 
%
%
A minimal set of rotations with the sign factors, $R^{(\sigma_R)}$, 
contains all the information needed to express the WFM in terms of $\mathbf{l}$. 
In this way, Turov obtained invariants describing weak ferromagnetism for all crystal classes \cite{Turov2010}.

For example, tetragonal fluorides MnF$_2$ and CoF$_2$ (space group $P4_2/mnm$) show a two-sublattice AFM state with $l\|c$. 
Ruthenium dioxide, RuO$_2$, has the same crystal symmetry and was recently suggested to have the same AFM order \cite{Berlijn2017,Zhu2019}.
In contrast to MnF$_2$ and CoF$_2$, RuO$_2$ is a metal, which is why it attracted many theoretical and experimental studies \cite{Smejkal2020,Smejkal2022a,Feng2022,Liao2024,Bai2023}, although there is a growing evidence that this material remains paramagnetic at all temperatures \cite{Mukuda1999,Kessler2024,Smolyanyuk2024,Liu2024,Kiefer2025,Plouff2025}. 
Screw rotations, $\tilde{4}_{z}$ and $\tilde{2}_{x}$, interchange magnetic sublattices, which makes 
the Zeeman coupling, $-\lambda(H_x l_y + H_y l_x)$, in the free energy density $f$ invariant under all symmetry operations. 
The magnetization, $\mathbf{M} = - \frac{\partial f}{\partial \mathbf{H}} = \lambda(l_y,l_x,0)$, can be used to select one out of four AFM domains with an in-plane $\mathbf{l}$.
Table~\ref{tab:effects} also shows a few non-linear coupling terms of third order in $l$.

To all orders in $l$, the state with an out-of-plane N\'eel vector has zero $\mathbf{M}$, making it seemingly impossible to select a single domain.
However, an increase of $H\|c$ above $\sim 9.3$ T, at which MnF$_2$ undergoes the spin-flop transition into a state with an in-plane $\mathbf{l}$, and its subsequent decrease result in a single-domain state selected by a nonzero WFM in the spin-flopped phase \cite{Felcher1996}.  
A single AFM domain was also selected by cooling through the AFM phase transition and using various weak interactions: the non-linear magnetic susceptibility, described by the $l_z H_x H_y H_z$ term in $f$  \cite{Kharchenko2005}, the $\mathbf{l}$-dependent absorption of linearly polarized light resulting from the dielectric permittivity tensor, $\delta \varepsilon_{xy} = \delta \varepsilon_{yx} \propto l_z H_z$ \cite{Higuchi2016}, and the piezomagnetic coupling, $H_z l_z u_{xy}$, where $u_{ij}$ is a strain tensor   \cite{Baruchel1988}.

The A-type AFM ordering in CrSb \cite{Takei1963}
and $\alpha$-MnTe  \cite{Kunitomi1964}
(space group $P6_3/mmc$), is described by $3_{z}^{(+)}, \tilde{2}_{z}^{(-)}$, and $2_{x}^{(+)}$ with the Cartesian $x$ axis along the hexagonal [2$\bar{1}\bar{1}$0] direction. 
This symmetry does not allow for a  linear coupling between the N\'eel vector and magnetic field, the lowest-order terms being $\left(H_x 2 l_x l_y + H_y (l_x^2 - l_y^2)\right)l_z$ and  $H_z (3l_x^2 l_y - l_y^3)$. 
The last term gives rise to an out-of-plane magnetization, $M_z \propto l^3 \sin^3\psi$, in $\alpha$-MnTe with the in-plane N\'eel vector, $\mathbf{l} = l(\cos\psi,\sin\psi,0)$.
It makes possible to select 3  out of 6 domains, favored by the magnetic in-plane anisotropy, $K \cos^2 3 \psi$ with $K > 0$, by an out-of-plane magnetic field.
%

The selection of 3 domains is sufficient to maximize the AHE \cite{Gonzales2023,Kluczyk2024}: e.g. the expansion of domains with $\psi = \frac{\pi}{2},-\frac{\pi}{6},-\frac{5\pi}{6}$ and contraction of domains with $\psi = -\frac{\pi}{2},\frac{5\pi}{6}, \frac{\pi}{6}$ under $H\|c$, affects $\sigma_{xy} - \sigma_{yx} \propto \sin (3\psi)$.
The X-ray magnetic circular dichroism combined with the magnetic linear dichroism and photoemission electron microscopy has been used to image discrete $Z_6$-vortices with all 6 domains merging at the vortex core \cite{Amin2024}.  
Under an applied magnetic field $H\|c$, the domains with 3 phases shrink and the $Z_6$ vortices transform into $Z_3$ vortices. 
A similar transformation was observed in multiferroic hexagonal manganites under an applied electric field \cite{Choi2010}, resulting from a nonlinear coupling between the out-of-plane electric polarization and trimerization of the crystal lattice \cite{Artyukhin2014}.

Antiferromagnetically ordered spins in CrSb with $l\|c$ are perfectly antiparallel.
It might be possible to select a single-domain state using the nonlinear magnetic susceptibility,  described by $l_z\left(3 H_x^2H_y - H_y^3\right)$, or the piezomagnetic effect, $l_z\left[2 H_x u_{xy} + H_y (u_{xx} - u_{yy}) \right]$.


In the absence of spin-orbit coupling, the projection of the electron spin $\mathbf{s}$ on the N\'eel vector, $m = \left(\mathbf{s}\cdot\mathbf{l}\right)$, is conserved.
The spin splitting of electron bands is described by
\begin{equation}
\delta \varepsilon_{\mathbf{k}m} = \left(\mathbf{l}\cdot\mathbf{s}\right)g(\mathbf{k}),
\label{eq:spolrutile}
\end{equation}
where $\mathbf{k}$ is the wave vector of the electron. 
This form ensures invariance under arbitrary rotations of both $\mathbf{l}$ and $\mathbf{s}$  in spin space. 

Under a crystal symmetry operation, $R$, $\left(\mathbf{l}\cdot\mathbf{s}\right)$ is multiplied by $\sigma_R$. 
Invariance of $\delta \varepsilon_{\mathbf{k}m}$ implies $g(\mathbf{k}) \xrightarrow{R} \sigma_R g(R\mathbf{k})$, where $R\mathbf{k}$ is the electron wave vector after the transformation.
$\tilde{4}_z^{(-)}, \tilde{2}_x^{(-)}$ and $I^{(+)}$ symmetries of the antiferromagnetically ordered RuO$_2$ give  $g(k_x,k_y,k_z) = -g(-k_y,k_x,k_z) = - g(k_x,-k_y,-k_z) = g(-k_x,-k_y,-k_z)$. 
The microscopic time-reversal invariance requires $g(-\mathbf{k}) = g(\mathbf{k})$, independent of whether the crystal lattice has inversion symmetry or not. 
The  spin splitting near $\Gamma$-point is given by $g(\mathbf{k}) \propto k_x k_y$ ($d$-wave) \cite{Yuan2020,Gonzales2021}. 
The $d$-wave splitting in RuO$_2$ is a matter of controversy \cite{Liu2024,Fedchenko2024}.

Similar analysis shows that the spin splitting in CrSb and $\alpha$-MnTe near $\Gamma$-point is described by the $g$-wave, $Y_{4,+3}(\hat{\mathbf{k}})+Y_{4,-3}(\hat{\mathbf{k}})$:
\begin{equation}
\delta\varepsilon_{\mathbf{k}m} = A \left(\mathbf{l}\cdot\mathbf{s}\right) k_z \left(3 k_x^2 k_y - k_y^3\right),
\end{equation}
in agreement with ARPES measurements on CrSb \cite{Ding2024,Reimers2024} and MnTe \cite{Lee2024,Krempasky2024,Osumi2024} (although, in none of these experiments a single-domain state was selected).


Similarly to conducting ferromagnets, the spin-split electron bands in altermagnets can lead to an electrical generation of the spin current, $\mathbf{j}_i = (j_i^x,j_i^y,j_i^z)$, where the upper index is the spin polarization and $i=x,y,z$ denotes the current direction.
Invariance under rotations in spin space implies that in the absence of spin-orbit coupling, the spin current linear in the N\'eel vector can only be polarized along $\mathbf{l}$:
\begin{equation}
\mathbf{j}_i = \mathbf{l} \alpha_{ik} E_k,
\label{eq:spin-current}
\end{equation} 
where $\mathbf{E}$ is an applied electric field. 
Discussion of symmetry properties of Eq.(\ref{eq:spin-current}) simplifies, if one assumes that a crystal rotation $R$ only acts on lower indices and does not change the spin current polarization and N\'eel vector, since their rotation can be undone by a counter rotation in spin space. 
However, the right-hand side of the equation acquires an extra sign factor $\sigma_R$. 
For $\tilde{4}_z^{(-)}$ and $\tilde{2}_x^{(-)}$ symmetries of antiferromagnets with  the rutile crystal structure, only $\alpha_{xy} = \alpha_{yx}$ can be nonzero.  
The inverse effect -- the non-relativistic $\mathbf{l}$-dependent conversion of the spin current into the charge current -- is described by  $(j_x,j_y) \propto \left( \left( \mathbf{l} \cdot \mathbf{j}_y \right), \left( \mathbf{l} \cdot \mathbf{j}_x \right) \right)$. 

Spin current is even under time reversal, whereas the right-hand side of Eq.(\ref{eq:spin-current}) is odd, $\mathbf{l} \xrightarrow{T} - \mathbf{l}$, i.e. this $T$-odd SHE involves dissipation. 
The `spin-splitter current' originates from a non-equilibrium occupation of spin-polarized electron bands created by the charge current  \cite{Gonzales2021}.
The correction to electron velocity is: $\Delta \mathbf{v}_{\mathbf{k}m} = \frac{1}{\hbar} \frac{\partial \delta\varepsilon_{\mathbf{k}m}}{\partial \mathbf{k}} \propto \left(\mathbf{l}\cdot\mathbf{s}\right) (k_y,k_x,0)$. 
Under the applied electric field $\mathbf{E}$, $\langle \mathbf{k} \rangle \propto \mathbf{E}$,  and the spin current is given by  $j^a_x \propto \langle s^a {\rm v}_x \rangle = A l^a E_y$ and $j^a_y = A l^a E_x$. 
In this heuristic way, the fourth column in Table~\ref{tab:effects}, showing expressions for the $T$-odd SHE, can be derived from the spin splitting near $\Gamma$-point (third column)  for all materials listed in the table, including MnSb and MnTe, for which the $g$-wave spin splitting leads to a non-linear dependence of the $T$-odd spin current on the electric field. 
This non-linear SHE becomes a linear one under strains or magnetic fields, since the substitution of  $E_i E_j$ by $u_{ij}$ or $H_i H_j$ does not change symmetry properties, e.g. $\mathbf{j}_z \propto \mathbf{l} \left(2u_{xy}E_x + (u_{xx} - u_{yy}) E_y\right)$ is allowed by symmetry. 
The $T$-odd SHE in RuO$_2$ thin films is a matter of debate   \cite{Bai2022,Bose2022,Karube2022,Plouff2025} and the non-nonlinear SHE effect has not been reported. 

\begin{table}[htbp]
	\centering
	\begin{tabular}{|c|c|c|c|c|}
		\hline
		Material & $R^{(\sigma_R)}$ & WFM & Spin Splitting & T-odd SHE\\ 
		[0.4ex]
		\hline
		\(
		\begin{array}{c}
		{\rm MnF}_2\\
		{\rm RuO}_2
		\end{array}
		\)
		& $\tilde{4}_{z}^{(-)}, \tilde{2}_{x}^{(-)}$ & \(
		\begin{array}{c}
		H_x l_y + H_y l_x,  \left(\mathbf{l}\cdot\mathbf{H}\right)l_x l_y\\ 
		H_z l_x l_y l_z  \\
		\left(l_x H_y - l_y H_x\right)\left(l_x^2 - l_y^2\right)
		\end{array} \) & $\mathbf{l}k_xk_y$&
		\(\begin{array}{c}
		\left(\mathbf{j}_x,\mathbf{j}_y, \mathbf{j}_z\right)\\
		= A \mathbf{l}\left(E_y,E_x,0\right)
		\end{array}\)\\
		[0.4ex]
		\hline
		\(
		\begin{array}{c}
		{\rm CrSb}\\
		{\rm MnTe}
		\end{array}
		\)
		& $3_{z}^{(+)}, \tilde{2}_{z}^{(-)}$, $2_{x}^{(+)}$ & \(\begin{array}{c}H_x 2 l_x l_y l_z + H_y (l_x^2 - l_y^2)l_z\\H_z (3l_x^2 l_y - l_y^3)\end{array}\) & $\mathbf{l} k_z \left(3 k_x^2 k_y - k_y^3\right)$ &  
		\(\begin{array}{rcl} 
		\mathbf{j}_x &\!\!=\!\!& 2A \mathbf{l}E_xE_yE_z\\
		\mathbf{j}_y &\!\!=\!\!& A \mathbf{l} (E_x^2-E_y^2)E_z\\ 
		\mathbf{j}_z &\!\!=\!\!& B \mathbf{l} (3E_x^2E_y-E_y^3)
		\end{array}\)\\
		[0.4ex]
		\hline
		\(
		\begin{array}{c}
		{\rm Mn}_3{\rm Ir}\\
		{\rm Mn}_3{\rm GaN}
		\end{array}
		\)& \(
		\begin{array}{c}
		{\rm cubic}\\
		3_{[111]}, 2_{[001]}, 2_{[110]}
		\end{array}
		\) 
		& $H_x S_1^x + H_y S_2^y + H_z S_3^z$ & 
		$\mathbf{S}_1 k_x^2 + \mathbf{S}_2 k_y^2 + \mathbf{S}_3 k_z^2$
		& 
		\(\begin{array}{c}
		\left(\mathbf{j}_x,\mathbf{j}_y, \mathbf{j}_z\right)=\\
		A\left(\mathbf{S}_1 E_x,\mathbf{S}_2 E_y, \mathbf{S}_3 E_z\right)
		\end{array}\)
		\\ 
		[0.4ex]
		\hline
		\(
		\begin{array}{c}
		{\rm Mn}_3{\rm Sn}\\
		{\rm Mn}_3{\rm Ge}
		\end{array}
		\)
		&  \(
		\begin{array}{c}
		{\rm hexagonal}\\
		3_{z}, \tilde{2}_{z}, 2_{x}
		\end{array}
		\)  & 
		\(\begin{array}{c}
		H_1 \xi_1 + H_2 \xi_2 + H_3 \xi_3\\
		H_n = \left(\mathbf{H}\cdot\mathbf{f}_n\right)\\
		\xi_n = \left(\mathbf{S}_n\cdot\mathbf{f}_n\right)
		\end{array}\)
		& 	\(\begin{array}{c}
		\mathbf{S}_1 k_1^2 + \mathbf{S}_2 k_2^2 + \mathbf{S}_3 k_3^2\\
		k_n = \left(\mathbf{k}\cdot\mathbf{f}_n\right)
		\end{array}\)&
		\(\begin{array}{c}
		\mathbf{j}_i = A \sum_n \mathbf{S}_n (f_n)_i E_n\\
		E_n = \left(\mathbf{E} \cdot\mathbf{f}_n\right)
		\end{array}\)
		\\
		[0.4ex]
		\hline
	\end{tabular}
	\caption{Macroscopic effects in collinear and non-collinear altermagnets expressed in terms of order parameters. For hexagonal Mn$_3$Sn and Mn$_3$Ge, $\mathbf{f}_1 = \left(-\frac{\sqrt{3}}{2}, - \frac{1}{2}, 0\right)$,  $\mathbf{f}_2 = \left(+\frac{\sqrt{3}}{2}, - \frac{1}{2},0\right)$, $\mathbf{f}_3 = (0,1,0)$ in the Cartesian basis with the $x$ and $z$ axes parallel to the hexagonal [$\bar{1}\bar{1}$20] and [0001] directions, respectively.}
	\label{tab:effects}
\end{table}

\begin{figure}[h!]
\includegraphics[width=.4\textwidth]{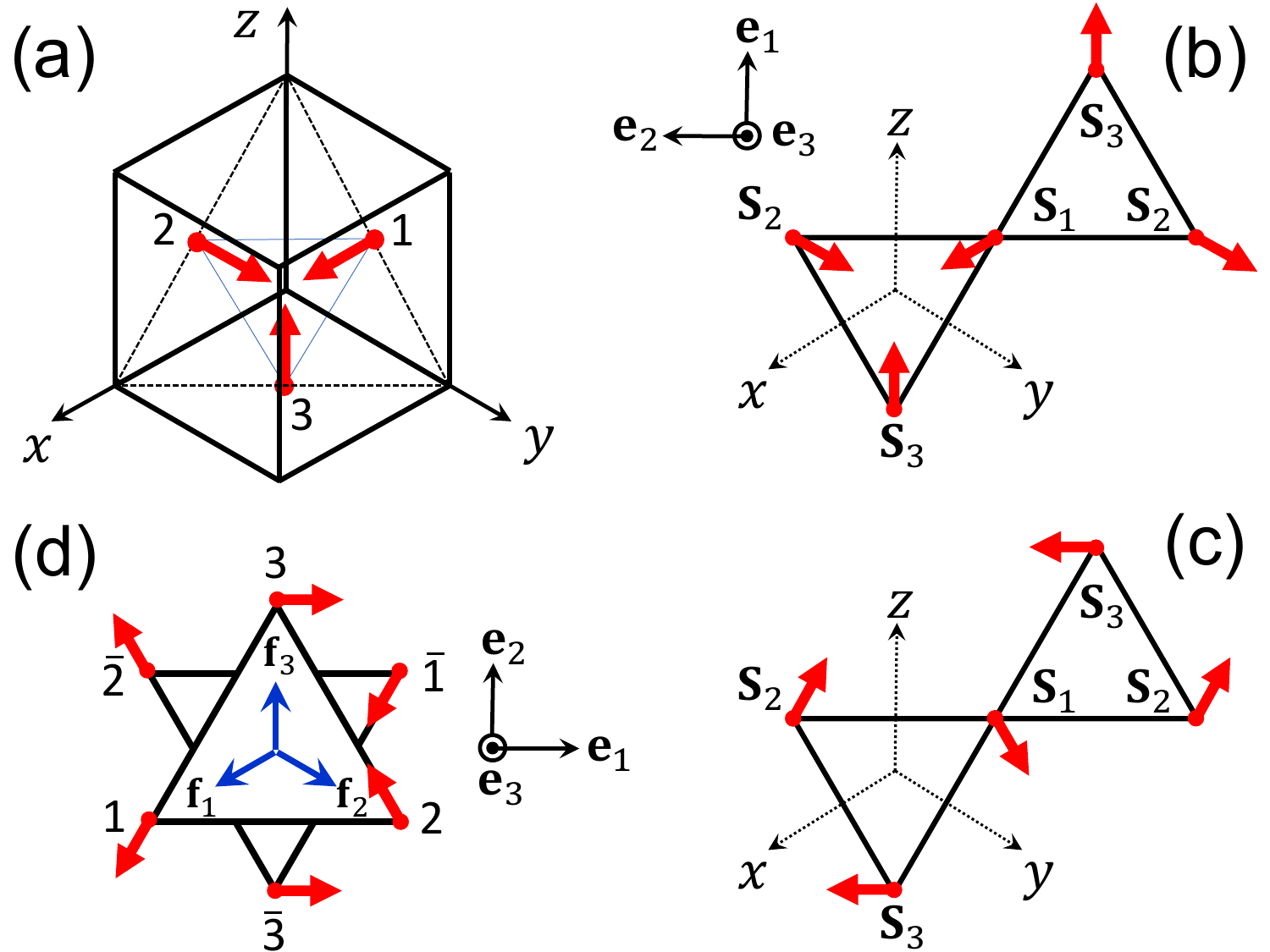}
\caption{(a) Non-collinear 120$^{\circ}$ spin ordering in the cubic Mn$_3$Ir. 
(b) Kagome spin lattice in the (111) plane. 
$\mathbf{e}_1 = \frac{1}{\sqrt{6}}\left(\bar{1},\bar{1},2\right), 
\mathbf{e}_2 = \frac{1}{\sqrt{2}}\left(1,\bar{1},0\right)$ and $\mathbf{e}_3 = \frac{1}{\sqrt{3}}\left(1,1,1\right)$. 
(c) Spin ordering in MnGaN. 
(d) Spin ordering in Mn$_3$Sn. Sites $\bar{1},\bar{2}$ and $\bar{3}$ are related to sites $1,2$ and $3$ by inversion. 
The unit vectors of the Cartesian frame: $\mathbf{e}_1, \mathbf{e}_2$ and $\mathbf{e}_3$, are along the hexagonal $[\bar{1}\bar{1}20], [1\bar{1}20]$ and $[0001]$ directions, respectively. 
In that frame,
$\mathbf{f}_1 = \left(-\frac{\sqrt{3}}{2}, - \frac{1}{2}, 0\right)$,  $\mathbf{f}_2 = \left(+\frac{\sqrt{3}}{2}, - \frac{1}{2},0\right)$ and $\mathbf{f}_3 = (0,1,0)$.}
\label{fig:noncollinear}
\end{figure}

A similar symmetry analysis can be applied to non-collinear AFM conductors with a 120$^\circ$-ordering that does not break inversion: 
cubic Mn$_3$Ir and Mn$_3$GaN (space group $Pm\bar{3}m$), 
and hexagonal Mn$_3$Sn and Mn$_3$Ge (space group $P6_3/mmc$). 
In (111) layers of the cubic materials and $ab$ layers of the hexagonal magnets, Mn spins form a Kagome lattice (see Fig.~\ref{fig:noncollinear}). 
The cubic unit cell contains 3 magnetic Mn ions and the absence of the spin-orbit coupling,
\begin{equation}
\mathbf{S}_1 + \mathbf{S}_2 + \mathbf{S}_3 = 0.
\end{equation}
The unit cell of hexagonal materials contains 6 magnetic ions at sites 1,2,3, forming a triangle in one Kagome layer, and sites $\bar{1},\bar{2},\bar{3}$, obtained by inversion, that form a triangle in a neighboring Kagome layer [see Fig.~\ref{fig:noncollinear}(d)]. 
Since the altermagnetic ordering does not break inversion, $\mathbf{S}_{\bar{n}} = \mathbf{S}_{n}$, all symmetry operations can be described using the three spins in one layer.

Under a rotation $R$: $S_n^a \rightarrow R^{ab} S_{R^{-1}n}^b$, where  $R^{ab}$ is a rotation matrix and $R^{-1}n$ is the magnetic site that transforms into $n$ under $R$, e.g. $\mathbf{S}_1 = (S_1^x,S_1^y,S_1^z)  \xrightarrow{3_{[111]}} (S_3^z,S_3^x,S_3^y)$ in Mn$_3$Ir.
The term, $H^x S_1^x + H^y S_2^y + H^z S_3^z$, describing weak ferromagnetism, is invariant under  $3_{[111]}, 2_{[001]}$ and $2_{[110]}$ rotations, which together with inversion and translations (not broken by the magnetic ordering) form generators of $Pm\bar{3}m$ space group. 
Similarly, the spin texture in reciprocal space near $\Gamma$-point (in non-collinear magnets spin is not conserved) is described by $\langle \mathbf{s} \rangle  \propto \mathbf{S}_1 k_x^2 + \mathbf{S}_2 k_y^2 +  \mathbf{S}_3 k_z^2$. 

Spin textures in non-collinear magnets have been predicted to give rise to an effective charge-to-spin conversion  \cite{Zelezny2017} and large tunneling magnetoresistance \cite{Shao2024}.
The `non-relativistic' spin current, $\mathbf{j}_x = A \mathbf{S}_1 E_x, \mathbf{j}_y = A \mathbf{S}_2 E_y$ and $\mathbf{j}_z = A \mathbf{S}_3 E_z$, flows in the direction of the applied electric field and its polarization lies in the plane spanned by the ordered spins. 

These results can be written in the form that also holds for Mn$_3$Sn  and Mn$_3$Ge. 
The crystal lattice of the hexagonal Kagome antiferromagnets also has a three-fold symmetry axis, a parallel two-fold (screw) axis and a perpendicular two-fold axis: $3_{[001]}, \tilde{2}_{[001]}$, $2_{[100]}$ (directions are given in hexagonal coordinates). 
The invariant describing WFM is $\sum\limits_{n=1}^{n=3} \xi_n H_n$, where  
$\xi_n = \left(\mathbf{S}_n\cdot\mathbf{f}_n\right)$ and $H_n = \left(\mathbf{H}\cdot\mathbf{f}_n\right)$. 
For cubic materials, $\mathbf{f}_1 = \hat{\mathbf{x}}$, $\mathbf{f}_2 = \hat{\mathbf{y}}$ and $\mathbf{f}_3 = \hat{\mathbf{z}}$, whereas for hexagonal materials, $\mathbf{f}_n$ points from the center to the corresponding vertex of the triangle [see Fig.~\ref{fig:noncollinear}(d)].
Similarly, one can obtain expressions for the spin texture near $\Gamma$-point,   
$\left \langle \mathbf{s} \right \rangle \propto \sum\limits_{n=1}^{n=3} \mathbf{S}_n \left(\mathbf{k}\cdot\mathbf{f}_n\right)^2$, and the spin-splitter current, $ j_{i}^{a} \propto \sum\limits_{n=1}^{n=3} S_n^a (f_{n})_i \left(\mathbf{E}\cdot\mathbf{f}_n\right)$.

The meaning of these results, summarized in Table~\ref{tab:effects}, becomes more transparent, if one describes the $120^{\circ}$ spin ordering by two orthogonal vectors, $\mathbf{V}_1 = \frac{1}{3}\left(2 \mathbf{S}_3 - \mathbf{S}_1 - \mathbf{S}_2\right) = \mathbf{S}_3$ and  $\mathbf{V}_2 = \frac{1}{\sqrt{3}}\left(\mathbf{S}_1-\mathbf{S}_2\right)$. 
Their vector product is the vector chirality:
\begin{equation}
\mathbf{V}_3 = \mathbf{V}_1 \times \mathbf{V}_2\\ 
= 
\frac{2}{3\sqrt{3}} 
\left(
\mathbf{S}_1 \times \mathbf{S}_2
+ \mathbf{S}_2 \times \mathbf{S}_3
+ \mathbf{S}_3 \times \mathbf{S}_1
\right).
\end{equation}
%
The direction of $\mathbf{V}_3$ is determined by magnetic anisotropies, which in Mn$_3$Ir  favor body diagonals and together with time reversal give rise to 8 magnetic domains.

In the Cartesian frame 
[see Fig.\ref{fig:noncollinear}(b)], 
 $\mathbf{V}_1 = \cos \psi \mathbf{e}_1 +  \sin \psi \mathbf{e}_2$, $\mathbf{V}_2 =  \chi \left(-\sin \psi \mathbf{e}_1 +  \cos \psi \mathbf{e}_2\right)$ and $\mathbf{V}_3 = \chi \mathbf{e}_3$, where $\psi$ is the  rotation angle in the (111) plane and $\chi = +1(-1)$ is referred to as  positive(negative) chirality. 
For $\chi = + 1$, $\mathbf{M} \propto \cos \psi \mathbf{e}_3$, whereas for $\chi = - 1$, $\mathbf{M} \propto \cos \psi \mathbf{e}_1 - \sin \psi \mathbf{e}_2$ lies in the (111) plane and rotates in the direction opposite to that of $\mathbf{V}_1$ as the angle $\psi$ varies.  

The strong second-order magnetic anisotropy, $-K \left[(S_1^x)^2+(S_2^y)^2+(S_3^z)^2\right]$ with $K>0$,   favors $\chi = + 1$ and $\psi = 0 \,\,\mbox{or}\,\,\pi$ in Mn$_3$Ir \cite{Szunyogh2009}.
In Mn$_3$GaN [see Fig.~\ref{fig:noncollinear}(c)], the chirality of the 120$^\circ$-state is also positive, but $\psi = \pm\frac{\pi}{2}$ \cite{Bertaut1968} (hence, no weak ferromagnetism and AHE), whereas in Mn$_3$NiN the angle $\psi$ is temperature-dependent  \cite{Fruchart1971}. 
The AFM domain selection, required to measure the AHE and T-odd SHE in thin films \cite{Iwaki2020} and single crystals under an applied strain \cite{Zuniga2023}, is helped by strong piezomagnetic effect observed in the cubic materials \cite{Zhang2025,Boldrin2018}.
The spin texture near $\Gamma$-point,   
$
\langle \mathbf{s} \rangle = A
\left( 
\mathbf{V}_1  d_{3k_z^2-k^2} 
+ \mathbf{V}_2 d_{k_x^2-k_y^2}
\right)
$, 
where $d_{3k_z^2-k^2} = \frac{1}{\sqrt{3}}(3k_z^2 - k^2)$ and $d_{k_x^2-k_y^2} = k_x^2 - k_y^2$, does not generate Berry curvature in reciprocal space, since  the ordered spins are coplanar.

The WFM of hexagonal magnets is only nonzero for negative vector chirality, $\chi = -1$, which is the case for  Mn$_3$Sn and Mn$_3$Ge.
For  $\mathbf{V}_1 = (\cos\psi,\sin\psi,0)$, $\mathbf{M} \propto (1-\chi)\left(\cos \psi, - \sin \psi, 0\right)$ lies in the $ab$ plane and rotates in the direction opposite to that of  $\mathbf{V}_1$ \cite{Tomiyoshi1982}. 
The $z$-component of the WFM and Hall vectors proportional to third power of spins ordered in the $ab$ plane \cite{Jianpeng2017} is forbidden by symmetry.
The WFM makes possible to select one out 6 ordered states: $\psi = 0,\pi,\pm \frac{\pi}{3}, \pm \frac{2\pi}{3}$ for  Mn$_3$Sn, and  $\psi = \pm \frac{\pi}{6}, \pm \frac{\pi}{2},\pm \frac{5\pi}{6}$ for Mn$_3$Ge, by an applied magnetic field \cite{Nakatsuji2015}, which was imaged by the scanning thermal gradient microscopy \cite{Reichlova2019}.

In hexagonal magnets, $\langle \mathbf{s} \rangle = A\left[\mathbf{V}_1 \left(k_x^2-k_y^2\right) + \mathbf{V}_2 2 k_x k_y\right]$ and the T-odd SHE is given by
\begin{equation}
\left(
\begin{array}{c}
\mathbf{j}_x\\ \mathbf{j}_y
\end{array}
\right)
=
A
\left[
\begin{array}{rc}
-\mathbf{V}_1 & \mathbf{V}_2\\
\mathbf{V}_2 & \mathbf{V}_1\\
\end{array}
\right]
\left(
\begin{array}{c}
E_x\\ E_y
\end{array}
\right)
\end{equation}  
(see also Table~\ref{tab:effects}). 
Interestingly, symmetry also allows for a non-relativistic T-even SHE with spins polarized along the vector chirality:
\begin{equation}
\left\{
\begin{array}{lcr}
\mathbf{j}_x &=& -B \mathbf{V}_3 E_y,\\
\mathbf{j}_y &=&  B \mathbf{V}_3 E_x.
\end{array}
\right.
\end{equation}



Magnetic orders breaking inversion or translational symmetry can also show transport phenomena that do not require spin-orbit coupling.
%
%
%
%
A collinear antiferromagnet with symmetries $3_z^{(+)}, 2_{x}^{(-)}$ and $I^{(-)}$ can show a non-linear SHE:
\begin{equation}
\left(
\mathbf{j}_x,\mathbf{j}_y,\mathbf{j}_z
\right)
=
\mathbf{l}\, E_z 
\left(
A E_x,A E_y, B E_z
\right),
\label{eq:nonlinearSHE}
\end{equation}
where $A$ and $B$ are numerical coefficients. 
Invariance under rotations in spin space suggests its non-relativistic origin. 
The spin splitting of electron bands is forbidden: time reversal invariance requires $g(-\mathbf{k}) = g(\mathbf{k})$, whereas the sign change of $\mathbf{l}$ under inversion implies  $g(-\mathbf{k}) = -g(\mathbf{k})$.
Instead, the current flowing under an applied electric field $E\|z$ can induce a magnetization:
\begin{equation}
\mathbf{M} \propto \mathbf{l}\, E_z.
\label{eq:LME}
\end{equation}
The electric current induced by the second electric field is then  converted into a spin current.
The Mott insulator, Cr$_2$O$_3$, has the same symmetry and shows the linear LME described by Eq.(\ref{eq:LME}) originating   from the nonrelativistic Heisenberg exchange striction \cite{Mostovoy2010}. 
This mechanism is responsible for a peak in temperature dependence of the magnetoelectric coefficient.
A relation, $\mathbf{M} \propto \frac{\partial \mathbf{l}}{\partial z}$, also allowed by symmetry,  implies that an AFM domain wall normal to the $z$ axis carries a net magnetic moment $\propto \mathbf{l}$ induced by the non-relativistic mechanism.

%
Simultaneous presence of even and odd (under inversion) magnetic orders induces an electric polarization in insulators \cite{Mostovoy2024} and an antisymmetric spin splitting of electron bands in conductors.
Consider the 120$^{\circ}$ spin ordering on a triangular lattice \cite{Hayami2020}, which has a wave vector $\mathbf{Q} = (1/3,1/3)$ and can be described in terms of two orthogonal vectors, $\textbf{V}_1$ and $\textbf{V}_2$:  
$
\mathbf{S}_{n} =  
\mathbf{V}_1 \cos{\mathbf{Q}\cdot\mathbf{X}_n} + 
\mathbf{V}_2 \sin{\mathbf{Q}\cdot\mathbf{X}_n}.
$
 Although this magnetic order breaks translational symmetry, it has a vector chirality and gives rise to the spin splitting:
$
\langle \mathbf{s} \rangle = [\mathbf{V}_1 \times \mathbf{V}_2] g(\mathbf{k}).
$ 
Since $\mathbf{V}_1$ is even and $\mathbf{V}_2$ is odd under inversion, $ g(-\mathbf{k}) = - g(\mathbf{k})$. Invariance under $3_z$ and $2_y$ implies,  
$g(\mathbf{k}) \propto k_x^3 - 3 k_x k_y^2$, near $\Gamma$-point (f-wave).



Collinear and noncollinear altermagnets can show large spin splitting of electron bands in the absence of spin-orbit coupling and for light magnetic and ligand ions. 
Low magnetic anisotropy makes ordered spin states susceptible to external stimuli, which is important for applications in spintronics.
Symmetry analysis in terms of vector order parameters is ideally suited for altermagnets, as it can incorporate spin rotations in response to applied electromagnetic fields, strains and stresses.

It is a pleasure to acknowledge fruitful discussions with A. V. Kimel, K. Sundararajan and G. Chavez Ponce de Leon.

\end{document}